\documentclass[sigconf]{acmart}
\acmConference[Submitted to ICSE 2022]{The 44th International Conference on Software Engineering}{May 21–29, 2022}{Pittsburgh, PA, USA}
\usepackage[utf8]{inputenc}
\usepackage{tcolorbox}
\usepackage{colortbl,hhline}
\definecolor{light-gray}{gray}{0.75}
\newcommand{\reasearchgoal}{The goal of this work is to help software developers and security specialists in measuring npm supply chain weak link signals to prevent future supply chain attacks by empirically studying npm package metadata.}

\title{What are Weak Links in the npm Supply Chain?}

\makeatletter
\def\author@bx@sep{0pc}
\makeatother
\author{\large Nusrat Zahan$^1$, Thomas Zimmermann$^{2}$, Patrice Godefroid$^{2}$, Brendan Murphy$^{2}$, Chandra Maddila$^{2}$, Laurie Williams$^1$}

\affiliation{\vspace{3pt}
      \institution{\normalsize {}$^1$ North  Carolina State University,  Raleigh, NC \country{USA}}}

\email{[nzahan, lawilli3]@ncsu.edu}   

\affiliation{\vspace{3pt}
      \institution{\normalsize {}$^{2}$ Microsoft Research, Redmond, Washington \country{USA}}}
\email{[tzimmer, pg, bmurphy, chmaddil]@microsoft.com}

\copyrightyear{2022}
\acmYear{2022}
\setcopyright{acmcopyright}\acmConference[ICSE-SEIP '22]{44nd International Conference on Software Engineering: Software Engineering in Practice}{May 21--29, 2022}{Pittsburgh, PA, USA}
\acmBooktitle{44nd International Conference on Software Engineering: Software Engineering in Practice (ICSE-SEIP '22), May 21--29, 2022, Pittsburgh, PA, USA}
\acmPrice{15.00}
\acmDOI{10.1145/3510457.3513044}
\acmISBN{978-1-4503-9226-6/22/05}
\begin{document}
\begin{abstract}
Modern software development frequently uses third-party packages, raising the concern of supply chain security attacks. Many attackers target popular package managers, like npm, and their users with supply chain attacks.
In 2021 there was a 650\% year-on-year growth in security attacks by exploiting Open Source Software's supply chain. Proactive approaches are needed to predict package vulnerability to high-risk supply chain attacks. \textit{\reasearchgoal} 

In this paper, we analyzed the metadata of 1.63 million JavaScript npm packages. We propose six signals of security weaknesses in a software supply chain, such as the presence of install scripts, maintainer accounts associated with an expired email domain, and inactive packages with inactive maintainers. One of our case studies identified 11 malicious packages from the install scripts signal. We also found 2,818 maintainer email addresses associated with expired domains, allowing an attacker to hijack 8,494 packages by taking over the npm accounts. 
We obtained feedback on our weak link signals through a survey responded to by 470 npm package developers. The majority of the developers supported three out of our six proposed weak link signals. The developers also indicated that they would want to be notified about weak links signals before using third-party packages. Additionally, we discussed eight new signals suggested by package developers.

\end{abstract}
\maketitle
\renewcommand{\shortauthors}{N. Zahan et al.}

\section{Introduction}\label{sec:Intro}
Modern software development frequently uses third-party packages, raising the concern of supply chain security attacks. According to Snyk~\cite{Snyk_report}, 96\% of applications use third-party packages, and 80\% of the code in the software supply chain comes from third-party packages. The scope and scale of the expanding supply chain also come with high-security risks~\cite{zimmermann2019small,Sonatype_2021}. Large package managers, like npm, maintain a centralized repository where developers can access and add third-party packages to their dependency tree easily~\cite{zimmermann2019small,duan2020towards}. Unfortunately, attackers can also leverage the same features of npm and inject malicious package updates into a software supply chain.

A recent report from Sonatype shows supply chain attacks has increased $650\%$ in 2021 on top of year-over-year growth of 430\% in 2020~\cite{Sonatype_2021} where attackers injected malicious code into benign packages~\cite{eslint,Event-stream,1337qq-js,etc-shadow-files, Snyk_script,codecov}. An example of a sophisticated supply chain attack is the SolarWinds attack~\cite{Solarwinds_lesson}. SolarWinds is a proprietary cybersecurity monitoring solution~\cite{Snyk_report,Solar_winds_425} whose customers include 425 Fortune~500 companies and at least nine U.S. federal agencies~\cite{Solar_winds_425}. More than 100 companies and federal agencies were exposed to the breach~\cite{Solarwinds_cost}. The attacker gained access to customer networks, systems, and data through a malicious package update~\cite{Solar_winds_425}. An incident of this magnitude raises significant concerns about the consequences of supply chain attacks.

In supply chain attacks, instead of exploiting latent vulnerabilities in source code, attackers inject malware directly into benign code that is likely to be deployed by users~\cite{Sonatype_2021,ohm2020backstabber}. A common strategy is to target the most commonly used packages in a dependency chain to infect a maximum number of users~\cite{Sonatype_2021}. In that way, bad actors can execute an attack that will propagate throughout the supply chain. Thus, popular large registries like npm, which hosts 1.8 million JavaScript packages as of 2021, are a highly targeted malware distribution channel for attackers due to heavy growth and dependence on JavaScript packages~\cite{Supply_chain,ohm2020backstabber,Sonatype_2021,ferreira2021containing,gonzalez2021anomalicious}.
 
 One way attackers can develop such supply chain attacks is by following a data-driven attack strategy. For example, an attacker can collect and analyze package metadata from the package registry to find and exploit the weakest links (e.g., less secure module, maintainers) and then execute an attack in the targeted supply. The dynamic nature of such attacks challenges conventional detection methods~\cite{ohm2020backstabber,duan2020towards,gonzalez2021anomalicious,ferreira2021containing} because the attacks are new and spread fast without the victim knowing where bad actors planted the malware in the supply chain. Hence, practitioners need proactive approaches to identify warnings, events that predict package susceptibility at low cost measures to prevent future supply chain attacks.

 \textit\reasearchgoal 
 
As we know the saying- "\textit{A chain is only as strong as its weakest link}". A weak signal is an indicator of an event or change that may become significant in the future~\cite{thorleuchter2013weak} and can break the supply chain. %
In this study, \textbf{we define a signal as a weak link if the signal exposes a package to a higher risk of a supply chain attack and an attacker can exploit the signal to execute a supply chain attack}. %
We focus on the relationship between neutral and suspicious metadata in a supply chain. We propose six weak link signals for npm package dependencies. To that end, we perform an empirical study on 1.63 million npm packages' metadata to quantify and prioritize  weak link signals in the npm supply chain. Our work addresses the following research questions:

\begin{itemize}
    \item\textbf{RQ1 (Quantification)}: What type of weak link signals exists in npm package supply chain?

    \item\textbf{RQ2 (Agreement)}: How do practitioners perceive the proposed weak link signals in the npm supply chain?
\end{itemize}

This paper makes the following contributions: 
 \textbf{First.} Six proposed weak link signals, three of which are confirmed as strong signals from a survey completed by 470 npm package maintainers.
    \textbf{Second.} Eight new weak link signals suggested by survey participants.
    \textbf{Third.} A framework\footnote{Due to reduce malicious use of the scripts, please contact the authors for access to the framework.} to collect, categorize and analyze package metadata in npm registry to evaluate weak link signals. 

\section{Background and Related Work}
A software supply chain attack is a cyber-attack that aims to infect organizations and end-users by targeting less-secure components in the supply chain~\cite{Supply_chain}. The supply chain encompasses everything that goes into or affects code from development through CI/CD pipeline until production deployment. Package registries like npm play an essential role in automating the software supply chain. Unfortunately, increased automation comes with a higher security risk. Supply chain attacks are considered critical because of the increasing reliance on third-party packages as a direct and transitive dependency. A single package dramatically increases a system’s attack surface due to the “nested” nature of dependencies~\cite{zimmermann2019small}. Therefore, we need to predict and prevent such weak links before the malicious code is distributed in the supply chain. %

\subsection{Key Terminology}
\subsubsection{\textbf{npm}} is a platform for publishing and hosting JavaScript packages which is the largest ecosystem to date. npm has a CLI tool for publishing and installing packages, and separately it has an online repository to host all the package and their metadata. All npm packages contain a file, called \textbf{package.json} - a central place to configure and describe how to interact with and run a package, and npm used this data to manage a package installation or handle the project's dependencies~\cite{package_json}. %

\subsubsection{\textbf{Primary Stakeholders}} Here, we discussed the relevant stakeholder roles deeply connected with the supply chain that can benefit from our research. \textbf{Package Maintainers} are responsible for developing and maintaining packages. They may receive and review pull requests from contributors and have write access to make changes in different stages of package development. \textbf{Package Contributors} can view the source code and can suggest code changes, and package maintainers approve their changes. 
\textbf{Ecosystem Administrators} manage the package registry framework and are responsible for maintaining the whole software ecosystem. 
\subsection{Attack Vector}Here, we overview the different attack vectors an attacker may use to introduce a supply chain attack. 1) \textbf{Malicious package release}: An attacker may publish malicious packages and hence trick other users into installing or depending on such packages~\cite{duan2020towards,ohm2020backstabber}. 2) \textbf{Social Engineering}: An attacker may manipulate a maintainer to hand over sensitive information~\cite{zimmermann2019small}. 3) \textbf{Account Takeover}: An attacker may compromise the credentials of a maintainer to inject malicious code under the maintainer’s name. 4) \textbf{Ownership transfer}: An attacker can show enthusiasm to maintain popular abandoned packages and transfer the ownership of a package~\cite{ohm2020backstabber}. 5)\textbf{ Remote execution}: An attacker may target a package by compromising the third-party services used by that package.
\subsection{Research on Supply Chain Work}
Many prior works have leveraged past supply chain records to show how catastrophic a supply chain attack can be. Zimmermann et al.~\cite{zimmermann2019small} provided evidence that popular packages and highly active developers in npm may suffer from single points of failure due to large direct and transitive dependency. 
Ohm et al.’s study~\cite{ohm2020backstabber} investigated different supply chain attacks from 174 malicious packages and contributed an enriched dataset for future research on malicious package detection. Duan et al.~\cite{duan2020towards} identified 339 malicious packages from their proposed unsupervised learning framework. Gonzalez et al.~\cite{gonzalez2021anomalicious} used repository and commit metadata to detect malicious packages automatically.

Although security researchers in academia and industry are actively investigating attacks on registries and proposing solutions, these approaches seem to be based on specific instances of malicious attacks. They are especially  effective to prevent malicious code distribution. Recent attacks show evidence that out-of-the-box exploit strategies will appear again and again~\cite{Solarwinds_opensource,codecov}. Any ad-hoc solution is not enough to prevent an attack that we have not witnessed yet. A better approach is needed to embrace a proactive strategy that predicts package susceptibility as a potential threat and help package managers and package maintainers adopt the best security practice to stay ahead of the attackers. 
\section{RQ1: Weak link signals} 
In this section, we describe our weak link signals and present quantification of those signals to answer RQ1.
\textbf{We define a signal as a weak link if the signal exposes a package to a higher risk of a supply chain attack and an attacker can exploit the signal to execute a supply chain attack}. If a bad actor targets the supply chain, they are going to follow the path of least resistance~\cite{viega2001building}, for example, finding weakest links from publicly available package metadata without penetrating the whole source code. In this study, we followed an approach the attackers might take to find the weakest link in the npm registry. An accurate estimation of the impact on a target application may not be given from a single weak link signal but identifying multiple weak link signals from different metadata aiming at a common target is probably a hints that this target will be impacted in future. 

One of the critical challenges in identifying the weak link signals was selecting the suspicious candidate of metadata for further investigations. To overcome this, we took advantage of the prior reported supply chain attack patterns reported in literature~\cite{ohm2020backstabber,duan2020towards,garrett2019detecting} and the author's security domain expertise. Additionally, we discussed our observation with security specialists from GitHub working on npm supply chain security. %
Based on our discussion, we focused on metadata associated with human involvement (e.g., maintainer, contributor information) since many past attacks involved human actors as a weak link to the package~\cite{weak_credential,weak_credential_2,weak_credential_3,Event-stream,Eslint_attack}. Then we selected functional metadata (install scripts)~\cite{eslint,etc-shadow-files,Snyk_script,1337qq-js,garrett2019detecting,duan2020towards,Solarwinds_opensource,cc-cleaner,adverline,Not_petya} that was exploited by attacker mostly. We also measure the attack surface (e.g., dependency, dependents) to understand the underlying reason that might lead to an attack attempt.

\subsection{Data Process }\label{sec:Metadata}
\subsubsection{\textbf{Data Collection}}We  captured a snapshot of 1,630,101 package.json files on June 7, 2021 through the  npm public API\footnote{\url{https://replicate.npmjs.com/_all_docs?include_docs=true}}. Each JSON file contains metadata properties of an npm package. We used the package ID (package name and version together form a unique identifier called ``ID'') as the primary reference while storing corresponding metadata for each package.  As a root, package ID maps to all other metadata like dependency information, package last modified time, scripts, versions, license, repository, unpacked package size, a total number of files, maintainers and contributors names and email addresses. We used Python libraries (JSON,Numpy, Pandas, and Pyscopg2) to extract all the relevant metadata from each package.json file which we stored systematically in a PostgreSQL database. Each package.json file contained multiple versions history of package metadata against one unique ID. %
In this study, we only used the most recent version of the metadata to analyze our weak link indicators. Apart from metadata extraction, we analyzed the maintainer reach and package reach to understand the impact of a package and its maintainers in the npm registry. Below we explain how we queried and measured package and maintainer reach-- 
\newline\textit{Package Reach}: We measured the package reach to evaluate a package's popularity in terms of dependents and downloads. Hence, we considered two metrics:  (1) the number of packages that depend on a package (dependents ), which we computed from a set of all the packages that have a direct dependency on a package; and (2) the number of downloads of a package in the past 12 months which we collected from public npm API\footnote{\url{https://api.npmjs.org/downloads/point/{period}[/{package}]}}.
\newline\textit{Maintainer Reach:} We measured the number of unique dependents that depend on  a maintainer's packages to evaluate the maintainer reach in npm. We queried against npm packages to compute maintainer reach, a set of unique packages where atleast one package of the maintainer is listed as a package dependency.
\subsubsection{\textbf{Exclusion Criteria}} \label{Exclusion}
We removed packages that might introduce noise (for example, wrong ownership, invalid metadata, etc.). In this study, we removed a package if the package has no \textbf{dependents %
and it meets with any one of the following IF} conditions:

\textbf{First}. If a package is tagged as a \textbf{security holding package}~\cite{Security_holding} %
and is removed from the \textit{npm} registry by the npm security team due to malicious activities. %
   Hence, the JSON file we received for such packages consisted of dummy metadata filled by the npm team. We removed 8,344 packages as security holding packages.  The ``descriptions'' and ``dis-tags'' property of the package.json file define the security holding status.
    
\textbf{Second}. If a package is \textbf{deprecated}~\cite{Deprecated}, 
 meaning, the package is not actively maintained by the maintainers. The package.json file contains a separate property called ``deprecated'' to indicate package deprecated status, which is assigned by the npm administrators or the maintainers themselves. Though a deprecated package does not always indicate an unusable package, npm and maintainers recommend using alternative packages or versions. %
While the deprecated packages are not actively maintained, end-users might still use them directly or transitively. Hence, we only removed deprecated packages in the recent version if no other packages in the npm registry used them in their dependency tree.  We removed 37,917 deprecated packages that used by none.
    
\textbf{Third}. If a package has \textbf{no repository and no license}. A repository is essential to track, organize, and validate the source coderights  and a license allows the OSS community to reuse the code. A package without a license indicates that the authors retain all source code ~\cite{package_json} and no repository is attached to verify otherwise. We have identified packages where both repository and license property was null or filled with invalid values (e.g., ``UNLICENSED'', ``XYZ'', ``personal use'' etc.). Hence, we only removed the packages from further considerations if they meets all three conditions: 1) no dependent \textbf{and}; 2) no license \textbf{and}; 3) no repository. We found 89,893 such packages and removed them from the database.

In total we removed \textbf{9\%(135,996)} packages that fits into our proposed exclusion criteria, and our final dataset contained \textbf{1,494,105 packages}, which were used for further analysis.

\subsection{Weak Link Signals} We briefly discuss six weak link signals proposed in this study, their attack models, and specific data analysis in this section. 
 
\subsubsection{\textbf{W1: Expired Maintainer Domain}} \label{W1} \textbf{An attacker can hijack a component if a maintainer's domain is expired and does not have 2FA authentication set up on their account}.
In general, any domain name can be purchased from a domain registrar allowing the purchaser to connect to an email hosting service to get a personal email address. An attacker can hijack a user's domain to take over an account associated with that email address. Typically, a domain hijacking attack occurs by 1) gaining unauthorized access to the registrar or 2) gaining access to the owner’s email address and then resetting the password. Domain hijacking is not a new notion. We have seen many attacks in the past, such as $perl.com$ hijacking~\cite{perl_domain}, where the attacker changed the domain registrar and renewed the domain expiration date until 2029, and then changed the DNS address. %

In the npm registry, an attacker can execute a more simplistic approach to hijack an email address. An attacker can track the domain of a maintainer in the domain registrar site. If the domain is expired and available for sale, the attacker can register and alter the DNS ``mail exchange'' (MX) records to hijack the maintainer's email address. In most cases, maintainer accounts are associated with an email address in the package registry. One could reset a npm account directly by email address unless the maintainer activated 2FA authentication or used different email address in user account. %

\textbf{Analysis of npm maintainers domain:}  To collect the maintainer email address, we extracted and stored the maintainer name and email address from  each package.json file. We then queried and split the domain name from each email address and counted the number of times a domain is used across all npm packages. Out of 93K domains in the 1.63M npm packages, $86\%$ were unique (appeared once), while the rest were from the public or organizational domain.
We hypothesize that all maintainer domains in the npm registry are up-to-date, and none of them are available for sale. To that end, we performed a bulk query of 93K domains in Godaddy %
a domain registrar site. We found 5,346 domains are available for sale. We picked a random sample of 50 domains to determine the true positive rate and checked each domain independently in Godaddy. We found 33 of them are not for sale. Due to the high false-positive rate, we manually verified 5,346 domains in Godaddy. \begin{tcolorbox}[colback=black!10!white,colframe=black!10!white,boxrule=0.0mm, boxsep=0.1mm]
We found 2,818 maintainer's domains are available for sale and can be purchased. These maintainers own 8,494 packages in the npm registry with average direct dependents of 2.43 packages and average downloads of 53K in the past 12 months.
\end{tcolorbox}We reported our findings to the npm security team to validate the email address for npm user accounts. We note that our bulk query may have a high false-negative rate since we found 47\% false positive from the 5,346 domain. Hence,npm may have more than 2,818 maintainers associated with expired domains %

\subsubsection{\textbf{W2: Installation Script:}} \label{W2}
\textbf{An attacker can use installation scripts to run commands that perform malicious acts through the package installation step~\cite{Snyk_best_practice}}. Install scripts run automatically either before, during, or after package installation when certain events are triggered. These scripts are used to make the installation process easy since they are automatically run by npm. However, for an attacker, such scripts create opportunities where ``sky is the limit''. The attacker could steal user-sensitive data or execute a new child process to create backdoor access or gain access to execute a series of commands remotely~\cite{eslint,etc-shadow-files,Snyk_script,1337qq-js,garrett2019detecting,duan2020towards}. Alternatively, the attacker can infiltrate the third-party dependence since that  installation script will run automatically by the targeted package and its users during installation. The rc, coa, ua-parser-js~\cite{coa_rc}, CCleaner~\cite{cc-cleaner}, Solarwinds~\cite{Solar_winds_425}, NotPetya~\cite{Not_petya}, and Adverline~\cite{adverline} data breaches are examples of such attacks where attackers targeted the remote server, third-party vendors to execute a large supply chain attack. %
Though the presence of installation scripts themselves are not a direct indication of maliciousness, the privilege to run automatically makes installation scripts a weak link signal in the supply chain. As best practices, even npm registry recommends avoiding install script- ``\textit{Don't use install...You should almost never have to explicitly set a preinstall or install script.}''\footnote{\url{https://docs.npmjs.com/cli/v7/using-npm/scripts}}.%
While evidence of such an attack through the installations script is not rare~\cite{eslint,etc-shadow-files,Snyk_script,1337qq-js,garrett2019detecting,duan2020towards,Solarwinds_opensource,cc-cleaner,adverline,Not_petya}, similar or perhaps even worse, attacks may happen in the future.

\textbf{Analysis of npm packages:} To collect script details, we extracted and stored the script key, which is the script’s name (e.g.preinstall) and the script value which contains the script path/shell commands from the package.json file. Then, we query and separate all the packages that have script keys like ``\%Install\%''. \begin{tcolorbox}[colback=black!10!white,colframe=black!10!white,boxrule=0.0mm, boxsep=0.1mm]
We found 2.2\% (33,249) of packages use install scripts, indicating that 97.8\% of packages may follow npm recommendation of not using the install script as best security practices. 
\end{tcolorbox} Additionally, we collected 3,635 malicious packages.json files from npm. They are similar to the security holding package.json file (see Section \ref{Exclusion}) with all dummy metadata. The only difference is that these JSON files have actual malicious script key and value pair, for example- the original directory of malicious code files or malicious shell script embedded in JSON files with other dummy metadata. To analyze the malicious JSON file, two researchers separately reviewed these scripts and compared results for verification. Since the scope of the project was limited to package.json files only, we were only able to analyze 485 (out of 3,635) packages where malicious shell command was embedded directly in the JSON file and 442 of them were in install scripts. From our analysis, we found four types of attack patterns that attackers use frequently-
\begin{itemize}
    \item \textbf{Transfer Users data} to third party server (e.g.- hostname, etc/shadow, /etc/passwd,/home/<user>/.ssh ). We found 344 packages that communicate and transfer data with third party server.
    \item \textbf{Download} malicious tool and run it to user machine. For example- download a crypto miner software and run it on user machine. We found 115 packages.
    \item \textbf{Reverse Shell} 12 packages opened reverse shell and transfer data to third party server.
    \item \textbf{Removed directories} 14 packages removed file/folder from current directories.
\end{itemize}\begin{tcolorbox}[colback=black!10!white,colframe=black!10!white,boxrule=0.0mm, boxsep=0.1mm]
We found 93.9\% (3,412) of malicious packages had atleast one install scripts, indicating that malicious attackers use install scripts frequently.
\end{tcolorbox}

\subsubsection{\textbf{W3: Unmaintained Package}} \label{W3}
\textbf{Attackers can target packages that are more likely to take over and sneak in malware due to lack of maintenance}. Differentiating between unmaintained and feature-complete packages that require no further releases is complex and ambiguous~\cite{vaidya2019security}. Even if the package may not require any maintenance by itself, it may need maintenance due to security issues in its dependencies or use new syntax to improve performance, bug fixing \& documentation improvements. In 2020, the average time to remediate security issues was 68 days in open source projects~\cite{snyk_report_mttu} and 66\% of security vulnerabilities in npm packages remain unpatched~\cite{Sonatype_2020}. Hence, the time required to remediate a security issue in unmaintained packages remains unknown.
We considered unmaintained packages from two directions: 1) Inactive packages; and 2) Inactive maintainers.
\smallskip

\textbf{Inactive Packages:} %
We considered a package inactive if the package's last modification time in the package.json file is past two years.
Other prior work~\cite{vaidya2019security,Sonatype_2020} has defined inactivity as one year gap. However, many npm packages have low complexity with a few lines of code and may not require any recent update.  Thus, we consider two years as an inactivity period.

An attacker can exploit vulnerabilities in all applications that directly or transitively depend on vulnerable code as long as the vulnerable dependency or the package itself remains unfixed~\cite{zimmermann2019small}. The response time to fix such issues is undetermined for inactive packages, and end-users may remain infected due to unawareness of the security threat. A deprecated package (see Section \ref{Exclusion}) exposes even higher security risk since the maintainers no longer maintain the package to fix security issues. An attacker can inject malware in widely used but unmaintained packages. A prominent example of such an attack is the ``Mailparser'' attack~\cite{MailParser}, an old deprecated package. An attacker indirectly reaches the ``MailParser'' through a relatively new package named ``getcookies'', which had indirectly been made into the nested dependency chain of MailParser. %
\smallskip

\textbf{Inactive Maintainers:} We defined an inactive maintainer if the maintainer had no active packages in the past two years. An attacker can target packages with inactive maintainer(s) because any attack will remain undetected due to the inactivity of maintainers. Therefore, distinguishing between active and inactive maintainers is necessary to identify the packages that require maintenance. Although inactive maintainer's packages are a subset of inactive packages, we must identify them separately. Because an inactive package may have maintainers who are active in other packages, whereas inactive maintainers indicate the maintainer is inactive in the entire npm registry. %

\textbf{Analysis of npm packages:} We extracted and stored the ``time'' property of the package.json file to measure the number of packages that have been inactive for the past two years. We identified inactive maintainers by evaluating the last modified time properties for all packages corresponding to an individual maintainer. A package where none of the maintainers are active elsewhere in the entire package registry is determined as inactive maintainers of unmaintained packages. We also considered deprecated packages as unmaintained since they are unmaintained officially by the maintainer. We separated the deprecated package where the last modification time passed our threshold value because the deprecation was declared later. %
\begin{tcolorbox}[colback=black!10!white,colframe=black!10!white,boxrule=0.0mm, boxsep=0.1mm]
We found 58.7\% of packages and 44.3\% of maintainers are inactive in the npm registry. There are 5,532 additional deprecated packages where the deprecation date passed our threshold value.
\end{tcolorbox}
The package.json does not provide individual maintainer activities history. Hence, distinguishing inactive maintainers from active packages was not plausible.

\subsubsection{\textbf{W4: Too many Maintainers}} \label{W4}
\textbf{A package with too many maintainers will provide an attacker many targets to exploit account takeover and social engineering attacks.} Having too many maintainers requires security assurance of many maintainers. The open nature of open source software and npm not enforcing the two-factor authentication make it harder to ensure the maintainer's secure account. Bad actors may leverage the possible oversight of a large team and hide their identity by compromising one maintainer profile or performing social engineering to access the package. The fewer number of maintainers facilitates better security in terms of better communication.  

Our hypothesis is supported by previous work of Meneely et al.~\cite{meneely2009secure} where they empirically showed projects with more developers has more vulnerabilities. Zimmermann et al.~\cite{zimmermann2019small} also addressed our concern where they suggested that the value of over 20 maintainers in a npm package is questionable. %
A recent attack on inactive rc and coa ~\cite{coa_rc} packages shows that additional or inactive maintainers pose a security threat. In both packages, malicious code was injected via the compromised maintainer's account in November. Since the attack, the current version has fewer maintainers than before. Though we do not know how the attacker compromised the maintainer account, we have observed the change in maintainer metadata after the attack was resolved.

\textbf{Analysis of npm packages:} Though the exact number of maintainers  varies from project to project, in case of npm the number of maintainers is more likely to be less because npm is building upon reusing small JavaScript libraries. 1.5 million npm packages have an average cost of 1.7 maintainers, which makes sense as small packages may not require many maintainers. Hence, analyzing the whole data set is not practical to identify the unusual package with too many maintainers. We picked the top 1\% (14,941) packages in npm ranked by the total number of maintainers.  We extracted and stored the list of maintainers corresponding to each package in a SQL database and then ranked them by the total number of maintainers in the package. %
\begin{tcolorbox}[colback=black!10!white,colframe=black!10!white,boxrule=0.0mm, boxsep=0.1mm]
We found that our selected 1\% packages had an average of 32.4 maintainers per package, which was 19 times more than average package maintainers in the entire registry.
\end{tcolorbox} %

\subsubsection{\textbf{W5: Too many contributors}} \label{W5}
\textbf{An attacker can sneak in malicious code, bypassing the maintainer's radar when a maintainer is responsible for many contributors.} Many prior research shows that when multiple contributors change a file, the file is more likely to have more failures~\cite{bird2011don,meneely2009secure,nagappan2008influence,bird2009does} which may include security issues. These observations motivated our next signal: A maintainer-to-contributors ratio or increased number of contributors increases the security risks. 

Contributors may vary in knowledge, skill, and experience. Package quality will inevitably suffer if the maintainers do not pay enough attention to review the pull request from contributors. Especially where an average of 1.7  maintainers maintain JavaScript packages, many contributors bring additional responsibility for maintainers in product functionality or security. If the maintainers do not pay enough attention, contributors with malicious purposes can include potential backdoors into code, and malicious code will merge. An attacker can target packages with many contributors where the attacker can do social engineering to become a trusted contributor and make some minor contribution to gain trust and then sneak in malicious code. %

\textbf{Analysis of npm packages:} We extracted and stored the contributor(s) list from package.json files. We measured the ratio between the total number of maintainers to the total number of contributors of corresponding packages. %
Out of 1.5 million npm packages, only 2.6\% (38,913) of the packages have listed contributors, and the average maintainer to contributors ratio is 3:2. %
To understand the extreme cases where package maintainers added many contributors, we picked the top 1\% packages in npm where the maintainer to contributor ratio was minimum. \begin{tcolorbox}[colback=black!10!white,colframe=black!10!white,boxrule=0.0mm, boxsep=0.1mm]
We found that the selected 1\% (389) packages had an average maintainer to contributor's ratio of 1:40, which was 60 times more than the average ratio.
\end{tcolorbox} 
\subsubsection{\textbf{W6: Overloaded Maintainer}} \label{W6}
\textbf{An attacker may target a maintainer who owns many packages because the maintainer may not have enough time to maintain security of all the packages}. Overloaded maintainers are weak links if the maintainers 1) have a large number of dependents; 2) use a large dependency chain in packages, attackers may inject malware in their dependency; or 3) if they have many inactive packages, an attacker may try to take over those packages. A well-known attack on an overloading maintainer is event-stream attack~\cite{Event-stream}, where an original maintainer handed over a popular npm package ownership to a malicious maintainer simply because the package was inactive and the original maintainer does not need that package anymore. We queried the same maintainer in our database. We found that he is an overloaded maintainer, and he owns 62\% inactive packages with more than 30k direct dependents. Although one may argue that a maintainer who owns many packages may consider as a sign of stability over any new maintainer. while we agree with the statement, we want to include such a maintainer to the supply chain administrator's security radar. Attackers are more likely to target such maintainers if the maintainer is overloaded and does not have enough time to maintain all of his packages equally. We propose the supply chain administrator should follow up the security measure of overloading maintainers such as minimum package dependency, ownership transfer of inactive packages, two-factor authorization set up on their account, active domain registration.

\textbf{Analysis of overloaded maintainers:} We analyzed the overloaded maintainer using our maintainer reach metric~(section \ref{sec:Metadata}). We found that 48.2\% maintainers own more than one package, whereas only 24.8\% %
have downstream users. We again picked the top 1\%(4,743) maintainers ranked by higher maintainer reach to understand the extreme cases. \begin{tcolorbox}[colback=black!10!white,colframe=black!10!white,boxrule=0.0mm, boxsep=0.1mm]
We found that the top 1\% maintainers own an average number of 180.3 packages with direct dependents of 4,010 average packages
\end{tcolorbox} We analyzed further to understand risk factors associated with inactive packages and downstream dependencies. We found that 30\% (1,442) of maintainers do not hold any inactive packages, %
however, 70\% shows otherwise. In terms of package dependency, we found that 80\% of packages have dependencies, which indicates that overloading maintainers are responsible for their dependency chain security to protect the downstream users.

\section{Case Study}
This section provides three case studies of our proposed weak link signals.
First, we present our analysis in popular packages. We hypothesize that the popular package maintainers are aware of such weak links and will avoid them to protect the downstream user's security. Second, we present a case study on malicious packages identified by our signals, and at the end, we present an example of a data-driven supply chain attack.

\subsection{Case study \#1: Popular packages} Considering 650\% growth in supply chain attacks as an indication~\cite{Sonatype_2021}, attackers will likely continue to target popular packages and maintainers as a preferred path to exploit downstream victims at scale. Hence, we analyzed whether weak link signals exist in popular packages.%

To measure package popularity, we used the package reach metrics from Section {\ref{sec:Metadata}.} We picked the top 10,000 packages from 1) package dependents and 2) package downloads analysis; and combined the two lists, removing duplicates, into a combined popular package sample. The sample included 14,892 packages (1\% of total packages) as popular with an average of 937.4 dependents and 88.5 million downloads in 12 months. We note that the popular package analysis does not consider transitive dependents. Hence, the impact of a weak link in a popular package may present a higher supply chain risk than presented.

\smallskip
\textbf{Expired Domain (W1)} Among the popular packages, \textbf{33} packages (average direct dependents: 382.9 and average downloads: 11.1 million) have at least one maintainer with an expired domain. These accounts can be used to compromise the package unless two-factor authentication is enabled. %
\begin{tcolorbox}[colback=black!10!white,colframe=black!10!white,boxrule=0.0mm, boxsep=0.1mm]
W1: 12,637 packages that depend on popular packages were exposed to higher supply chain risk due to expired domains.
\end{tcolorbox}

\textbf{Install Scripts (W2)}: Among the popular packages, \textbf{362} packages (average direct dependents: 1,416.3 and downloads average: 34.6 million) have install scripts. This is an encouraging result, showing that 97.5\% of popular packages are aware of the risks and avoid using install scripts. %
\begin{tcolorbox}[colback=black!10!white,colframe=black!10!white,boxrule=0.0mm, boxsep=0.1mm]
W2: 362 popular packages had install scripts and exposed 1,416 packages  on average to attacks through install scripts.
\end{tcolorbox}
    
\textbf{Unmaintained Package (W3)}: Of the popular packages, \textbf{38\% (5,645)} packages (average direct dependents: 422.4 and downloads average: 76.1 million) had an inactive status. Interestingly, 560 of them (average direct downstream dependents: 1369.3 and downloads average: 32.8 million) were deprecated. Deprecated packages (Section~\ref{Exclusion}) are a classic example of unmaintained packages where maintainers officially declare the package unmaintained. %
We also found 619 inactive maintainers who still own 645 popular packages. %
\begin{tcolorbox}[colback=black!10!white,colframe=black!10!white,boxrule=0.0mm, boxsep=0.1mm]
W3: 38\% of popular packages were inactive, and 560 of them were deprecated. 
645 popular packages did not have any active maintainers. An inactive package exposed 422 packages on average to higher supply chain risk.
\end{tcolorbox}

\textbf{Too many Maintainers (W4)}: Among the popular packages, \textbf{421} packages (average direct dependents: 269.2 and downloads average: 41.6 million) had an average of 34.6 maintainers. 

Large packages may need more maintainers. Hence we hypothesize that 421 popular packages are large compared to the other popular packages. To test the hypothesis, we extracted and stored the ``unpacked size'' and ``file count'' property to measure the package size. We created two samples: 1) 14,471 popular packages without 421 packages (average of 54 files, 696.8 MB unpacked size, and 2.4 maintainers) and 2) 14,892 popular packages (average of 55 files, 697.8 MB unpacked size, and 3.4 maintainers). We found the 421 packages added an average cost of one maintainer without any significant difference between package size and file counts. Hence, having too many maintainers in a JavaScript package does not necessarily indicate packages are large. %
\begin{tcolorbox}[colback=black!10!white,colframe=black!10!white,boxrule=0.0mm, boxsep=0.1mm]
W4: 421 popular packages had an average of 34.6 maintainers and potentially exposed their dependents to higher supply chain risk.
\end{tcolorbox}
    
\textbf{Too many contributors (W5)}: Among the popular packages, \textbf{23} packages (average direct dependents: 6458.04 and downloads average: 64.2 million) had an average maintainer to contributor's ratio of 1:37. Therefore, on average, one maintainer has to manage 37 contributors' to secure packages against malicious code and malicious contributors.
\begin{tcolorbox}[colback=black!10!white,colframe=black!10!white,boxrule=0.0mm, boxsep=0.1mm]
W5: 23 popular packages are exposed to higher supply chain risk due to  maintainers having to manage an average of 37 contributors.
\end{tcolorbox}

\textbf{Overloaded Maintainers (W6)}: Among the popular packages,  \textbf{9,871 (66.3\%)} are owned by popular maintainers who manage many packages (average direct dependents: 1039.02 and downloads average: 110.3 million). Attackers can try to compromise these maintainers account to exploit downstream victims at scale. %
\begin{tcolorbox}[colback=black!10!white,colframe=black!10!white,boxrule=0.0mm, boxsep=0.1mm]
W6: 2,491 overloaded maintainers own 9,871 popular packages.
\end{tcolorbox}
\subsection{Case study \#2: Installation scripts} \label{case study 2}

Even though the focus was not on detecting malicious packages, our analysis of installation scripts revealed malicious activity within packages. Some packages have shell commands embedded directly in the package.json file, including malicious scripts. We generated a keyword lists from our 485 malicious packages analysis (section \ref{W2}) and queried those keywords in 2.2\% packages where they had install scripts. %

We identified 74 packages where the installation scripts included keywords such as- curl, wget, /etc/shadow, /etc/passwd. Of those, 11 were found to be malicious packages and rest were benign. All the malicious packages performed DNS lookups and send user-sensitive data to a specific URL. Three of the malicious package included shell commands for both windows and UNIX OS. 

After three months of our initial data collection, we collected a new security holding package list from npm to validate our result. Our findings aligned with the npm security specialists; they identified 10 out of the 11 packages as malicious. We reported the remaining malicious package to the npm security team.

This analysis demonstrates the risk of having a dependency on packages that include installation scripts.

\subsection{Case study \#3: Data-Driven Attacker} 

\begin{figure}[!htb]%
\centering \includegraphics[width=3.16in]{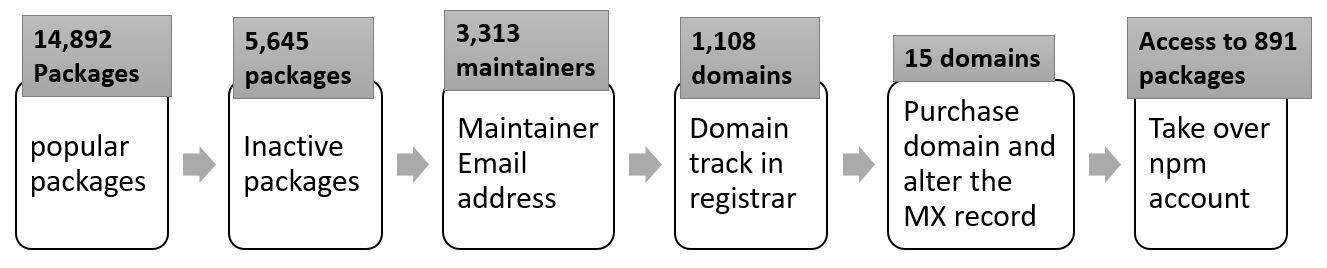} \caption{Data-driven Attack by combining W1 and W3}
\label{fig:case_study_3} 
\end{figure}

To illustrate we used the sample of popular packages (14,892) sample for this case study. Figure \ref{fig:case_study_3} shows an example of a data-driven attack that combines two of the signals: Expired Maintainer Domain~(W1) and Unmaintained packages~(W3). An attacker can scan the npm registry for popular packages and download relevant metadata from the package.json file. Then an attacker can easily extract unmaintained packages (5,645) and maintainer email addresses (1,108) from package metadata. The next step would be looking for domain availability in the domain registrar site (e.g., GoDaddy). %

In this stage, we have identified 15 domains available for sale in Godaddy. An attacker can purchase these domains and alter the MX record to hijack the maintainer's email addresses (section \ref{W1}). In general, npm requires an email address to set up an user account; attackers can reset npm accounts by using those 15 email addresses to access 899 npm packages. %

Another example of a data-driven attack would be looking for overloaded (W6) and inactive maintainers (W3) in popular packages. An attacker can perform social engineering to take over these packages. We have found 25 popular packages associated with 16 overloaded inactive maintainers. Out of 25, for three packages (average package dependents 117 and downloads 6,906) the last update year was 2013.
\begin{table}[!t] \renewcommand{\arraystretch}{1.3} \caption{A subset of combination signals quantified from our popular package analysis} \label{tab:data_driven}
\centering
\begin{tabular}{ p{70pt} || p{25pt} ||p{70pt} || p{25pt} }
 Signal combination & package count & Signal combination & package count\\
\hline\hline
\hline\hline
{W6~$\cap$~W3~$\cap$ W4} & 32 & {W2~$\cap$~W3~} & 86\\\hline
{W6~$\cap$~W3~$\cap$ W1} & 5 & {W6~$\cap$~W1~}& 17\\\hline
{W6~$\cap$~W3~$\cap$W2} & 38 & {W6~$\cap$~W2~}& 187\\\hline
{W3~$\cap$~W4} & 32 & {W6~$\cap$~W3~}& 3356\\\hline
\end{tabular}
\end{table}
Table~\ref{tab:data_driven} shows a subset of other possible combinations and how they would reduce a large number of 1M+ npm packages to a small number of candidate packages for potential supply chain attacks.

\section{RQ2: Survey} \label{survey}
In this section, we discuss our RQ2:\textit{How do practitioners perceive the proposed weak links in the npm supply chain?} To that end, we conducted a survey on npm package maintainers. We selected the top 10\% (47,433) of the maintainers ranked by the number of owned packages as survey candidates. We chose this selection criterion for the following two reasons: The maintainers are 1) experienced with JavaScript packages since they own many packages in npm, and 2) part of a large supply chain as many packages use these maintainer's packages as dependencies. %

Our survey is designed to capture practitioners' views on our proposed weak links and whether they want to be notified if any of the proposed weak links exist in their package dependency graph.  

Table \ref{tab:survey} is a complete summary of our survey. We attached the agreement and disagreement of practitioners regarding each weak link signal in the second and third columns. The last column provides the percentage of the practitioner who wants to be notified about such signals.  We observed that package maintainers supported three of our proposed signals as a weak link, and they would want to be notified about these signals existence in the package dependency graph.

\textbf{Agreement:} Out of six signals, more than 50\% of practitioners supported W1~(Expired Maintainer Domain), W2~(Install Scrips) and W3~(Unmaintained Packages) as a weak link signal. We observed that practitioners indicated a desire to be notified of weak link signals despite not supporting the weak link signal directly. As shown in Table \ref{tab:survey}, the percentage in the "want to be notified" column exceeds the percentage of ''Agreed as a weak link" for W2, and W3. %

\textbf{Disagreement} Our survey findings show that practitioners did not support W4~(Total Number of Maintainers), W5~(Maintainer to contributors ratio) and W6~(Packages per maintainer) as weak link signals. More than 40\% of people disagreed with these signals, whereas less than 20\% supported this as a weak link. To understand the context behind why practitioners think otherwise, we analyzed practitioner's comments in the open-ended questions. 
For example-- \textit{``Historically, I would have agreed with ``Too many maintainers'' being a risk, but as long as they are known people to you....I believe it to be ok.''} or \textit{``Many contributors is not a signal, it's a desired state of open source and if all are reviewed by a reliable maintainer, they are no risk.''} Both of the statements indirectly support our weak links assumptions under certain condition like ``reliable maintainer'' or ``as long as maintainer are known people''. %
Moreover, we do not claim that having any of these signals means a bad package; instead, our proposed metrics are guidelines for practitioners to make informed decisions about the use of the package.
\begin{table*}
\renewcommand{\arraystretch}{1.3} \caption{Survey responses from 470 npm package maintainers} \label{tab:survey}
\centering
\begin{tabular}{ p{250pt} ||| p{50pt} | p{50pt} |||  p{70pt} }
 Weak link Signal & Agreed as a weak link  & Disagreed as a weak link & Want to be notified\\
\hline\hline
{W1: A maintainer's email address is associated with an expired domain} & \cellcolor{light-gray}58.5\% (275) & 13.19\%(62) & \cellcolor{light-gray}55\% (258)\\\hline
{W2: A package has pre and post install scripts} & \cellcolor{light-gray}44.8\% (211) & 22.34\% (105) & \cellcolor{light-gray}57.4\% (270)
\\\hline
{W2: A package script has shell commands- CURL,WGET, NC,DIG} & \cellcolor{light-gray}67.45\% (317) & 8.09\% (38) & \cellcolor{light-gray}72.6\% (341)  \\\hline
{W3: A package has no update for X years} & \cellcolor{light-gray}58.7\% (276)  & 16.6\%(78) & \cellcolor{light-gray}63.6\% (299)\\\hline
{W3: A maintainer is inactive for X months/years} & \cellcolor{light-gray}57.7\% (271)  & 16.6\%(78) & \cellcolor{light-gray}63.6\% (299)\\\hline
{W4: A package has many maintainers} & 15.3\% (72) & \cellcolor{light-gray}54.7\% (257) & 11.7\% (55) \\\hline
{W5: A maintainer reviews a large number of pull requests from many contributors} & 17.87\% (84) & \cellcolor{light-gray}46.6\% (219) & 8\% (38) \\\hline
{W6: A maintainer owns too many packages} & 18.7\% (88) & \cellcolor{light-gray}49.4\% (232) & 9.6\% (45)\\\hline
\end{tabular}
\end{table*}

\textbf{New signals proposed by Maintainers} We asked an open-ended question for the respondents to recommend additional signals that we should consider in future work. Out of 470 practitioners, we received 213 responses for new signals. To label the new signals, two researchers separately reviewed the 213 responses and compared results. We included the new signal if the signal was raised by ``at least'' two respondents.  In some cases, the practitioner's intent was unclear, and the comment was discarded. We summarize the two most frequently mentioned concepts:

\textbf{Maintainers}: 41 practitioners in some way mentioned maintainers being a risk. We have identified the most frequent discussion on maintainers and propose the following four signals:
\begin{itemize}
    \item \textbf{Ownership transfer or adding new maintainers}: Any sudden change in a package maintainers list is proposed to be a weak link from practitioners. Practitioners would want to know about such changes in the package dependency graph if a package transfers the ownership or adds any new maintainers.
    \item\textbf{Maintainer Identity}: Practitioners commented on the role of maintainer expertise and identity verification in the supply chain. A maintainer with a real picture, organizational background, and email address, linked social media or repository, history of co-authoring with other maintainers will make a maintainer reliable over any new maintainers. Although npm provides the list of packages owned by maintainers, enforcing maintainers to add real identity or experience may be a big security improvement in the community.
    \item\textbf{Maintainer Two-Factor Authentication} A maintainer missing two factored authentication (2FA) for package hosting or releasing a new version or login to the npm account is a weak link. 2FA authentication should be enforced for all maintainers to publish a package.
\end{itemize}
\textbf{Integration of version control software}: 52 (24.4\%) of the responses were related to version control software(VCS), package repository and npm integration.
\begin{itemize}
    \item \textbf{No source code repository}: When a package has no or wrong public source code repository/homepage/VCS or the linked repository is archived, the access to review source code is restricted, forcing users to trust a package blindly.
    \item \textbf{npm package vs source code repository} The practitioners raise the concern to validate the published npm package against the code on the source code repository. Hence, all files inside a given package must match the exact contents in the repository.
    \item \textbf{CI/CD pipeline}: Missing CI/CD infrastructure to test code and build of npm packages. The practitioners also mentioned that the type of CI/CD services matters. Whether CI/CD service providers or self-hosted infrastructure, the practitioners prefer details on testing, code coverage, or alerts on the use of compromised CI/CD systems from past security incidents. %
    \item \textbf{Open pull request}: A package with many open issues and pull requests (PR) indicates a poorly maintained package. One can view if a package has open issues in npm online repository. However, practitioners commented on adding such information in the package dependency graph.
    \end{itemize}
Since our analysis is limited to package metadata from npm, we did not consider any repository-related weak link signal in this study but can be considered as a future research direction. 
\section{Limitations} We proposed and studied several weak link signals inferred from npm metadata and which can be used to evaluate security risks associated with npm packages. However, we do not claim that these weak link signals are the only ones that should be considered. Additional other signals suggested by practitioners indicate that further research is needed on weak link identifications. Another limitation of this work is that three of our six proposed signals W4~(Too many Maintainers), W5~(Too many Contributors), W6~(Overloaded Maintainer) were hard to empirically evaluate because we did not have enough metadata on maintainers activities to validate these in contrast to the clearer "ground truth" we have for W1, W2, W3. To address this limitation, future work could try to collect and leverage additional maintainer metadata, including commit history, vulnerability fixes, and maintainers turnover (how maintainers of a package are added or removed over time). Unfortunately, such metadata is not currently available in npm. Another limitation of our study is that we analyzed npm ecosystem only, which is the largest package manager ecosystem today, but we did not evaluate other package manager ecosystems which may have similar weak link signals. Despite these limitations, we believe there is an advantage of taking action to remediate such weak links instead of waiting for an attacker to explore them first. By performing a preliminary analysis on npm supply chain weak links, we hope to create an awareness within the community on why we need a risk model to mitigate such issues. %

\section{Discussion}
Through this study, we  increase awareness and visibility in detecting weak link signals to enhance supply chain security. Although this study does not provide a complete solution to mitigate the proposed weak link signals, we expect the findings will aid practitioners to predict package susceptibility in supply chain. The following subsections discuss our recommendations on how to strengthen supply chain security proactively instead of reacting to attacks.
 
 \subsection{Risk Model} %
  Currently, JavaScript npm package users have access to public data on package dependencies, as well as information about the package maintainers. However, npm packages do not have any overall "health" or security score. Therefore, estimating a security risk score associated with installing and using a new package is currently hard. Our work identifies several weak link signals which could be used for this purpose. We envision a community effort that could address this problem in the near future. For package managers, npm could compute and display a risk model based on weak link signals. Package managers would then know where their packages stand and improve their security scores by addressing the identified weaknesses. Such a risk model would allow package users to make more educated, data-driven decisions and comparisons before including new packages into their supply chains.  To this end, we suggest adding automated indicators for W1, W2, W3 in the OpenSSF Metrics \cite{OpenSSFMetrics} and OpenSSF Scorecard \cite{Scorecard} projects. 
 
\subsection{Control in Package Release}New packages are being released in npm by different maintainers every day. Within a timeline of five months, the npm has hosted more than 200K new packages, increasing the size and complexity of the npm supply chain. As a package managers, npm could validate any new release against the risk model that could be developed following the recommendations of this study. After a particular package is validated using the risk model, npm could publish the package and make it available to users. If the validation is unsatisfactory, npm could ask the maintainers of that package to improve its security by reducing weak link signals, for instance by confirming specific requirements impacting secure CI/CD pipelines for different OS environments or limiting the use of install scripts. 

\subsection{Trusted Package System} In our survey (Section~\ref{survey}), practitioners mentioned grading packages based on security risk, aligning with our proposed risk model above, which may be in conjunction with the OpenSSF Metrics \cite{OpenSSFMetrics}, Scorecard \cite{Scorecard}, and Best Practices Badge \cite{Badges} projects. Respondents indicated a recommendation system in terms of different security grades. We acknowledge that any kind of grading and measuring of such a large ecosystem is difficult and expensive. In that case, npm could prioritize or separate packages based on package reach in terms of dependents and downloads: the above security risk model could be implemented only for the most popular npm packages, which would then form a new "trusted package system". npm could exclude new packages or packages without any dependents or with few downloads from this trusted package system to avoid friction with the publication of new packages.

\section{Summary}In this work, we presented a framework that helps prioritization and quantification of weak link signals in the npm supply chain. %
We hope that identifying these weak link signals will help practitioners structure discussions and analyses of such issues. As part of an ongoing investigation, we submitted a list of suspicious packages to the npm security team to take necessary action, %
such as taking over packages from inactive maintainers, freezing the maintainer account if the maintainer domains are available for sale, or measuring security status of deprecated packages. Additionally, npm now requires mandatory 2FA on maintainers of top-100 npm packages by dependents~\cite{2fa}. Our second contribution consists of a list of new weak link signals proposed by a survey of npm practitioners.  We hope our work will promote further research on weak link signal identifications. Moreover, implementing a risk model around these signals would allow developers to make more educated, data-driven decisions before including packages into their software. 

\medskip
\textbf{ACKNOWLEDGMENTS:} We thank Bas Alberts and Max Schafer from GitHub for encouraging us to pursue this research and for their valuable feedback. We acknowledge the npm package maintainers contributions to our study.
We also thank the NCSU Realsearch group for valuable feedback.  In particular, we thank Aishwarya Seth and Parth Kanakiya for their assistance with the qualitative analysis of respondent open-ended responses and malicious scripts, respectively.  This work was funded by a Microsoft Research internship, and by Cisco and NCSU Secure Computing Institute.

\bibliographystyle{ACM-Reference-Format}
\bibliography{npm}
\end{document}